\documentclass[preprint,showpacs,preprintnumbers,amsmath,amssymb]{revtex4}

\usepackage{graphicx}
\usepackage{dcolumn}
\usepackage{bm}

\usepackage{feynmf}
\usepackage{slashed}

\begin{document}

\title{Unification of Gauge Couplings and the Higgs Mass in Vector-Like Particle Theories Extended into NMSSM}
\author{Yi-Lei Tang}
\affiliation
{Institute of Theoretical Physics, Chinese Academy of Sciences, \\
and State Key Laboratory of Theoretical Physics,\\
P. O. Box 2735, Beijing 100190, China}
 \email{tangyilei10@itp.ac.cn}

\date{\today}

\begin{abstract}

The minimal supersymmetry standard model (MSSM) extended with vector-like theories have long been discussed. If we extend the vector-like MSSM theory into NMSSM and let the vector-like particles couple with the singlet (S), we find out a natural way to generate the vector-like particle masses near 1TeV through the breaking of the $Z_3$ group. Compared with MSSM+vectorlike models, vector-like models extended into NMSSM contain more yukawa couplings and can help us adjust the renormalization group (RG) trajectories of the gauge couplings in order to unify the intersections. They can also help press down the gauge RG-$\beta$ functions for a $5+\overline{5}+10+\overline{10}$ model, in order for the RG-trajectories of the gauge couplings to unify before the Landau-pole. We also discuss the higgs mass contributed from the vector-like sectors in this case.

\end{abstract}
\pacs{12.60.Jv, 14.60.Hi, 14.65.Jk, 14.80.Da}

\keywords{supersymmetry, vector-like generation, LHC}

\maketitle
\section{Introduction}
Minimal supersymmetry standard model (MSSM) is a way to extend the standard model (SM) \cite{Primer}. Within this framework, every particle is paired up with a super-partner with a different spin. One of the features of this model is that it can also automatically unify the the gauge-coupling constants in the energy scale $\sim10^{16}$GeV \cite{SUSY_GUT}, as is required by grand unification theory (GUT). Another way to extend the SM is to add extra copies of the $U(1)_Y \times SU(2)_L \times SU(3)_C$ multiplets. In order to construct an anomaly-free theory, the simplest way is just to add extra SM generations \cite{4th1, 4th2, 4th3, 4th4}. Compared with the theories extended with chiral 4th generation, theories extended with vector-like (VL) generation is easier to survive among the experimental limits due to their particular vector-like mass parameters.

MSSM extended with VL generations have long been discussed \cite{VL, Martin:2009bg, Martin:2010dc, Liu:2009cc, Liu_With_Lu, Chang:2013eia, TopSeeSaw}. In order not to disturb the gauge coupling unification scale, only copies of $SU(5)$ $5+\overline{5}$, or $10+\overline{10}$ multiplets are the candidates to be added into the theory. Up to one-loop level, one $10$ multiplet modifies the renormalization group (RG) $\beta$ functions of the gauge coupling constants three times the same as one $5$ multiplet. If we would like the VL particles to be of the mass $1$TeV, and require the gauge couplings to meet with each other before they knock into Landau Pole, we can only choose $N_5$ copies of $5+\overline{5}$, where $N_5 \leq 3$, or one $10+\overline{10}$ \cite{GMSB_GUT}.

The situation of 4 copies of $5+\overline{5}$ theory, or $5+\overline{5}+10+\overline{10}$ theory is subtle. One-loop calculation unifies the gauge coupling constants with a value of 2-5, which approaches the Landau pole so much. Two-loop corrections usually contribute a positive value to the gauge RG-$\beta$ functions, thus directly accelerate the gauge RG-trajectories to blow up before they meet.

In MSSM, effective higgs mass receives extra loop contributions from the top/stop yukawa coupling \cite{Primer}. Higher order of stop mass raises the higgs mass while aggravates the tension of fine-tuning, and VL theories supply another source of the higgs mass. In this case, VL particles should directly couple with the SM-like higgs, and thus affect the Higgs phenomenology \cite{Elisabetta}.

As is well-known, MSSM suffers from the $\mu$ problem, and adding VL generations cannot solve this problem at all. The most economic way to solve this problem is to extend MSSM into NMSSM by adding a singlet (S) \cite{NMSSMInt}. The vacuum expectation value (vev) of the S naturally generates roughly $\mu^2 \sim B \mu$, and the super-potential term $\lambda S H_u H_d$ can also add up to the higgs quartic couplings, thus raise up the higgs mass.

If we extend the VL-MSSM with a singlet S, just like the NMSSM, we may take the advantages of both these theories. Similar consequences have been discussed in \cite{NMSSMVLAncester1, NMSSMVLAncester2, NMSSMVLAncester3}. Models with a scalar singlet and VL fermions without supersymmetry is also studied in \cite{HeHongjianS}. However, in our model, the VL-particles couple with S, so the VL-mass terms naturally come from the vev of the singlet Higgs rather than being inputed ``by hand''. By setting appropriate value to the yukawa constants near GUT-scale, the mass-spectrum of the VL-fermions can be partly predicted. Gauge coupling trajectories also receive extra contributions from the yukawa coupling constants, thus the unification can be improved by adjusting the values of the yukawa couplings in $5+\overline{5}$ and $10+\overline{10}$ models. In $5+\overline{5}+10+\overline{10}$ model, Extra yukawa coupling constants also contribute a non-ignorable minus value in two-loop gauge RG-$\beta$ functions, thus the gauge-coupling RG trajectories might meet before the perturbative theories lose effect.
\newpage	

\section{General Model}

If we extend the ordinary $Z_3$ NMSSM theory, we should assign the VL super-fields $Q$, $\bar{Q}$, $U$, $\bar{U}$, $D$, $\bar{D}$ with $Z_3$ charges in order for them to be coupled with the S. These $Z_3$ charges will also keep VL fermions massless before $Z_3$ breaks. Appropriate assignment will also forbid the VL particles to mix with the SM $Q_3$, $U_3$ fields, which is highly limited by experiments. The quantum numbers assigned to the VL particles are listed in Table \ref{Fields_Quantum_Number}. However, only $5+\overline{5}+10+\overline{10}$ model involves all the fields listed in Table \ref{Fields_Quantum_Number}. In our discussion about $5+\overline{5}$ model and $10+\overline{10}$ model, only part of the fields are needed. We should also note that all MSSM quarks and leptons are ignored except that the effects of the top sectors are considered in our discussions due to their large yukawa coupling constant.
\begin{table}
\begin{tabular}{|c|c|c|c|c|c|c|}
\hline
 & $U(1)_Y$ & $SU(2)_L$ & $SU(3)_C$ & $Z_3$ & R-Parity & Description\\
\hline
$Q$ & $\frac{1}{6}$ & $2$ & $3$ & $e^{i \frac{4 \pi}{3}}$ & $-$ & Vector-like quark doublet.\\
$\bar{Q}$ & $-\frac{1}{6}$ & $2$ & $\bar{3}$ & $1$ & $-$ & Vector-like anti-quark doublet.\\
$U$ & $-\frac{2}{3}$ & $1$ & $\bar{3}$ & $1$ & $-$ & Vector-like right-handed up-type quark.\\
$\bar{U}$ & $\frac{2}{3}$ & $1$ & $3$ & $e^{i \frac{4 \pi}{3}}$ & $-$ & Vector-like right-handed up-type anti-quark.\\
$D$ & $-\frac{1}{3}$ & $1$ & $\bar{3}$ & $1$ & $-$ & Vector-like right-handed down-type quark.\\
$\bar{D}$ & $\frac{1}{3}$ & $1$ & $3$ & $e^{i \frac{4 \pi}{3}}$ & $-$ & Vector-like right-handed down-type anti-quark.\\
\hline
$L$ & $-\frac{1}{2}$ & $2$ & $1$ & $e^{i \frac{4 \pi}{3}}$ & $-$ & Vector-like lepton doublet\\
$\bar{L}$ & $\frac{1}{2}$ & $2$ & $1$ & $1$ & $-$ & Vector-like anti-lepton doublet.\\
$E$ & $1$ & $1$ & $1$ & $1$ & $-$ & Vector-like right-handed electron\\
$\bar{E}$ & $-1$ & $1$ & $1$ & $e^{i \frac{4 \pi}{3}}$ & $-$ & Vector-like right-handed anti-electron.\\
\hline
$H_u$ & $\frac{1}{2}$ & $2$ & $1$ & $e^{i \frac{2 \pi}{3}}$ & $+$ & Up-type higgs doublet.\\
$H_d$ & $-\frac{1}{2}$ & $2$ & $1$ & $e^{i \frac{2 \pi}{3}}$ & $+$ & Down-type higgs doublet.\\
$S$ & $1$ & $1$ & $1$ & $e^{i \frac{2 \pi}{3}}$ & $+$ & NMSSM Singlino higgs.\\
\hline
$Q_3$ & $\frac{1}{6}$ & $2$ & $3$ & $1$ & $-$ & SM 3rd generation quark doublet.\\
$U_3$ & $-\frac{2}{3}$ & $1$ & $\bar{3}$ & $e^{i \frac{4 \pi}{3}}$ & $-$ & SM right-handed top.\\
\hline
\end{tabular}
\caption{Fields with Their Assigned Quantum Numbers}
\label{Fields_Quantum_Number}
\end{table}

Here we are going to take a short description about the basic NMSSM. The superpotential is \cite{NMSSMInt}
\begin{eqnarray}
W_{NMSSM}=\lambda H_u H_d S + \frac{1}{3} S^3 + y_t Q_3 H_u U_3,
\end{eqnarray}
and together we show the supersymmetry soft-breaking terms
\begin{eqnarray}
V_{NMSSM}^{soft}= m_{Hu}^2 |\tilde{H}_u|^2 + m_{Hd}^2 |\tilde{H}_d|^2 + M_S^2 |\tilde{S}|^2 + (\lambda A_{\lambda} \tilde{H}_u \tilde{H}_d \tilde{S} + \frac{1}{3} \kappa A_{\kappa} \tilde{S}^3 + y_t A_{y_t} \tilde{Q}_3 \tilde{H}_u \tilde{U}_3+h.c.).
\end{eqnarray}

The convention of the vacuum expectation values of the higgs fields is
\begin{eqnarray}
H_u^0 &=& v_u + \frac{H_{uR} + i H_{uI}}{\sqrt{2}} \nonumber \\
H_d^0 &=& v_d + \frac{H_{dR} + i H_{dI}}{\sqrt{2}} \\
S &=& v_s + \frac{S_R + i S_I }{\sqrt{2}}, \nonumber
\end{eqnarray}
so the MSSM-like superpotential term $\mu_{eff} H_u H_d= \lambda v_s H_u H_d$ is generated.

Since the VL fermions receive mass terms from $v_s$, their masses are actually in the same quantity as $v_s$ in most cases. In NMSSM, we usually concern about the $\mu_{eff}$ and $\lambda$, and then $v_s=\frac{\mu_{eff}}{\lambda}$. If $100\text{GeV} < \mu_{eff} < 1\text{TeV}$, $0.01<\lambda<1$, which is usually applied for successful electro-weak symmetry breaking, we can derive that $1\text{TeV}<v_s<100\text{TeV}$. Collider bounds on VL quarks have been reviewed in \cite{Earlier}, and the CMS Collaboration recently published their lower bound of VL top-like quark mass to a value of 687-782 GeV \cite{CMS_Limit}, so in our discussions below, we assume all of our VL fermions lie in the mass scale 1TeV, which is near the bound, although it is very easy to accumulate the VL mass towards $10 \sim 100$TeV by lowering $\lambda$ or raising $\mu_{eff}$. We also set 1TeV as the turning point in our RG-trajectory calculations.

\subsection{$5+\overline{5}$ Model}

The $5+\overline{5}$ model is the simplest model. It only contains vector-like down-type right-handed quark and anti-quark, $D$ and $\bar{D}$, together with vector-like leptonic doublet $L$ and $\bar{L}$. The VL particles can only couple with the $S$, as the superfield showed below,
\begin{eqnarray}
W_{5+\overline{5}}=\lambda_D \bar{D} D S + \lambda_L \bar{L} L S. \label{5p5W}
\end{eqnarray}
In literature, right-handed neutrino $N$ might be introduced so that $L H_u N$ vertices are discussed, however, here we ignore them.

The supersymmetry breaking soft terms should be added,
\begin{eqnarray}
V^{soft}_{5+\overline{5}}=m_D^2 (\tilde{D} \tilde{D}^{\dagger} + \tilde{\bar{D}} \tilde{\bar{D}}^{\dagger}) + m_L^2 (\tilde{L} \tilde{L}^{\dagger}+\tilde{\bar{L}} \tilde{\bar{L}}^{\dagger}) + (\lambda_D A_{\lambda_D} \tilde{\bar{D}} \tilde{D} \tilde{S} + \lambda_L A_{\lambda_L} \tilde{\bar{L}} \tilde{L} S + h.c.). \label{5p5V}
\end{eqnarray}
Notice that we have assumed that $\tilde{D}$, $\tilde{\bar{D}}$, or $\tilde{L}$, $\tilde{\bar{L}}$ share the same soft-mass term only for simplicity. We can observe from (\ref{5p5W}, \ref{5p5V}) that only singlet higgs sectors are involved here. However, as to be discussed below, this still contribute to the SM-like higgs mass.

\subsection{$10+\overline{10}$ Model}

The $10+\overline{10}$ model contains vector-like quark doublets $Q$, $\bar{Q}$, vector-like up-type quark singlets $U$ and $\bar{U}$, and vector-like electron singlet $E$ and $\bar{E}$. The yukawa coupling structure is richer than $5+\overline{5}$ theory due to the appearance of higgs doublets.
\begin{eqnarray}
W_{10+\overline{10}}=\lambda_Q \bar{Q} Q S + \lambda_U \bar{U} U S + \lambda_E \bar{E} E S + y_U Q H_u U + y_{\bar{U}} \bar{Q} H_d \bar{U}. \label{10p10W}
\end{eqnarray}
The corresponding soft-terms are listed below,
\begin{eqnarray}
V_{10+\overline{10}}^{soft}&=& m_Q^2 ( Q Q^{\dagger} + U U^{\dagger} ) + m_E^2 E E^{\dagger}
+ ( A_{\lambda_Q} \lambda_Q \tilde{\bar{Q}} \tilde{Q} \tilde{S} + A_{\lambda_U} \lambda_U \tilde{\bar{U}} \tilde{U} \tilde{S} \nonumber \\ 
&+& A_{\lambda_E} \lambda_E \tilde{\bar{E}} \tilde{E} \tilde{S} + A_{y_U} y_U \tilde{Q} \tilde{H_u} \tilde{U} + A_{y_{\bar{U}}} y_{\bar{U}} \tilde{\bar{Q}} \tilde{H_d} \tilde{\bar{U}} + h.c.) \label{10p10V}
\end{eqnarray}

\subsection{$5+\overline{5}+10+\overline{10}$ Model}

The $5+\overline{5}+10+\overline{10}$ model it not only a combination of $5+\overline{5}$ and $10+\overline{10}$, new terms also rise up.
\begin{eqnarray}
W_{5+\overline{5}+10+\overline{10}}=W_{5+\overline{5}}+W_{10+\overline{10}}+y_d Q H_d D + y_{\bar{d}} \bar{Q} H_u \bar{D} + y_L L H_d E + y_{\bar{L}} \bar{L} H_u \bar{E}.
\end{eqnarray}
The corresponding soft-terms are
\begin{eqnarray}
V^{soft}_{5+\overline{5}+10+\overline{10}}&=&V_{5+\overline{5}}+V_{10+\overline{10}}
+(A_{y_d} y_d \tilde{Q} \tilde{H_d} \tilde{D} + A_{y_{\bar{D}}} y_{\bar{d}} \tilde{\bar{Q}} \tilde{H_u} \tilde{\bar{D}} \nonumber \\
&+& A_{y_L} y_L \tilde{L} \tilde{H_d} \tilde{E} + A_{y_{\bar{L}}} y_{\bar{L}} \tilde{\bar{L}} \tilde{H_u} \tilde{\bar{E}}+h.c.).
\end{eqnarray}
During our discussions of the $5+\overline{5}+10+\overline{10}$ model, we would like to set $m_Q^2=m_D^2$ and $m_L^2=m_E^2$ for simplicity.

\section{$5+\overline{5}$ Model}

For the simplest $5+\overline{5}$ model, the extra vector-like particles couple with the S, thus contribute to the Higgs mass. It is much easier to calculate this contribution than the circumstances in $10+\overline{10}$, or $5+\overline{5}+10+\overline{10}$, because we only need to diagonalize $2 \times 2$ mass(-squared) matrices here. There are two down-type squarks, and the corresponding mass-squared matrix is
\begin{eqnarray}
\left[
\begin{array}{cc}
  m_D^2+\lambda_D^2 v_s^2 & v_s^2 \kappa \lambda_D - v_d v_u \lambda \lambda_D + A_{\lambda_D} \lambda_D v_s \\
  v_s^2 \kappa \lambda_D - v_u v_d \lambda \lambda_D + A_{\lambda_D} \lambda_D v_s & m_D^2 + \lambda_D^2 v_s^2
\end{array}
\right]. \label{SD_Matrix_55bar}
\end{eqnarray}
The mass matrix of the two slepton doublets is
\begin{eqnarray}
\left[
\begin{array}{cc}
  m_L^2+\lambda_L^2 v_s^2 & v_s^2 \kappa \lambda_L - v_d v_u \lambda \lambda_L + A_{\lambda_L} \lambda_L v_s \\
  v_s^2 \kappa \lambda_L - v_u v_d \lambda \lambda_L + A_{\lambda_L} \lambda_L v_s & m_L^2 + \lambda_L^2 v_s^2
\end{array}
\right]. \label{SL_Matrix_55bar}
\end{eqnarray}
Gauge D-terms are ignored, and so is done in our remaining sections. Notice that for each slepton doublet, the masses of charged slepton and the neutral slepton are degenerate in this model. To diagonalize (\ref{SD_Matrix_55bar}) and (\ref{SL_Matrix_55bar}) we acquire
\begin{eqnarray}
M_{\tilde{D}_1}^2 &=& m_D^2+\lambda_D^2 v_s^2 + v_s^2 \kappa \lambda_D - v_d v_u \lambda \lambda_D + A_{\lambda_D} \lambda_D v_s, \nonumber \\
M_{\tilde{D}_2}^2 &=& m_D^2+\lambda_D^2 v_s^2 - v_s^2 \kappa \lambda_D + v_d v_u \lambda \lambda_D - A_{\lambda_D} \lambda_D v_s, \nonumber \\
M_{\tilde{L}^C_1}^2=M_{\tilde{L}^N_1} &=& m_L^2+\lambda_L^2 v_s^2 + v_s^2 \kappa \lambda_L - v_d v_u \lambda \lambda_L + A_{\lambda_L} \lambda_L v_s, \label{M_SDL}\\
M_{\tilde{L}^C_2}^2=M_{\tilde{L}^N_2} &=& m_L^2+\lambda_L^2 v_s^2 - v_s^2 \kappa \lambda_L + v_d v_u \lambda \lambda_L - A_{\lambda_L} \lambda_L v_s, \nonumber
\end{eqnarray}
where $\tilde{D}_{1,2}$ are the two down-type squarks and $\tilde{L}^C_{1,2}$, $\tilde{L}^N_{1,2}$ indicate the two charged sleptons and the two neutral sleptons separately.

The masses of the down-type vector-like quark and the charged (neutral) lepton are
\begin{eqnarray}
M_D&=&\lambda_D v_s \nonumber \\
M_L^C=M_L^N&=&\lambda_L v_s. \label{M_DL}
\end{eqnarray}

Thus, we can take (\ref{M_SDL}) and (\ref{M_DL}) into Colemann-Weinberg potential under $\overline{MS}$ or $\overline{DR}$ scheme
\begin{eqnarray}
V_{CW}=\frac{1}{64 \pi^2} \left[ \sum_{scalars}{ m_s^4 N_s \left( \ln{\frac{m_s^2}{Q^2}} - \frac{3}{2} \right) }-\sum_{fermions}{ m_f^4 N_f \left( \ln{\frac{m_f^2}{Q^2}} - \frac{3}{2}\right) } \right], \label{CW_Potential}
\end{eqnarray}
where Q is the renormalization scale, and $N_s$, $N_f$ indicate the degrees of freedom of the particles. $N_s$ and $N_f$ take the value of 6 for colored fermionic or complex scalar particles, and 2 for colorless ones. Notice that the sum over fermions means to sum over all weyl-spinors, so each Dirac particle contributes an extra factor of 2 there.

If we assume that the SM-like Higgs mass-eigenstates to be in alignment with the vacuum expectation value (vev), that is to say, $\alpha=\frac{\pi}{2}-\beta$, where $\alpha$ is the mixing angle of the Higgs mass-eigenstates, the SM-like higgs mass should be added with a term
\begin{eqnarray}
\Delta m_h^2 &=& \frac{1}{2} \sin^2{\beta} \left( \frac{\partial^2 V_{CW}^{5+\overline{5}}}{\partial v_u^2}-\frac{1}{v_u} \frac{\partial V_{CW}^{5+\overline{5}}}{\partial v_u}\right) + \frac{1}{2} \cos^2{\beta} \left( \frac{\partial^2 V_{CW}^{5+\overline{5}}}{\partial v_d^2}-\frac{1}{v_d} \frac{\partial V_{CW}^{5+\overline{5}}}{\partial v_d}\right) \nonumber \\ 
&+& \sin{\beta} \cos{\beta} \frac{\partial^2 V_{CW}^{5+\overline{5}}}{\partial v_u \partial v_d} \nonumber \\
&=& \frac{3}{8 \pi^2} \sin^2{\beta} \cos^2{\beta} \lambda^2 \lambda_D^2 \left(\ln{\frac{M_{\tilde{D}_1}^2}{Q^2}} + \ln{\frac{M_{\tilde{D}_2}^2}{Q^2}}\right) \nonumber \\
&+& \frac{1}{4 \pi^2} \sin^2{\beta} \cos^2{\beta} \lambda^2 \lambda_L^2 \left(\ln{\frac{M_{\tilde{L}_1}^2}{Q^2}} + \ln{\frac{M_{\tilde{L}_2}^2}{Q^2}}\right). \label{Higgs_Mass_55}
\end{eqnarray}

It seems strange that the SM-like Higgs mass listed in (\ref{Higgs_Mass_55}) is Q-dependent, which is invisible in MSSM theories. In MSSM, the tree-level quartic coupling among Higgs fields only comes from the gauge D-terms, so loop contributions irrelevant to the gauge terms should not be renormalized in order not to break the gauge invariance. However, in the case of NMSSM, the appearance of $\lambda S H_u H_d$ also contribute to the Higgs quartic coupling, and receives the quantum correction from the field-strength renormalization constant $Z_S$ of S. We can then define $\lambda_{eff}=1+\frac{3}{64 \pi^2} \lambda_D^2 \left(\ln{\frac{M_{\tilde{D}_1}^2}{Q^2}} + \ln{\frac{M_{\tilde{D}_2}^2}{Q^2}}\right) + \frac{1}{32 \pi^2} \lambda_L^2 \left(\ln{\frac{M_{\tilde{L}_1}^2}{Q^2}} + \ln{\frac{M_{\tilde{L}_2}^2}{Q^2}}\right)$, and replace $\lambda$ with $\lambda_{eff}$ in the tree-level term $\lambda^2 v^2 \sin^2 2 \beta$, then we can also reach (\ref{Higgs_Mass_55}). Such kind of corrections have appeared in literature, e.g. \cite{LotsofMistakes}, although the method it used is too complicated to show the Q-dependence.

(\ref{M_SDL},\ref{M_DL}) through (\ref{CW_Potential}) also contribute to the CP-even singlet Higgs mass. Expand the consequence up to $\lambda_D^4$ and $\lambda_L^4$,
\begin{eqnarray}
\Delta m_S^2 &=& \frac{1}{2} \left( \frac{\partial^2 V_{CW}^{5+\overline{5}}}{\partial v_s^2} - \frac{1}{v_s} \frac{\partial V_{CW}^{5+\overline{5}}}{\partial v_s} \right) \nonumber \\
&=& \lambda_L^2 \frac{3 A_{\lambda_L} v_s^2 \kappa + 4 v_s^3 \kappa^2 + A_{\lambda_L} v^2 \sin{\beta} \cos{\beta} \lambda}{8 \pi^2 v_s} \ln{\frac{m_L^2}{Q^2}} \nonumber \\
&+& 3 \lambda_D^2 \frac{3 A_{\lambda_D} v_s^2 \kappa + 4 v_s^3 \kappa^2 + A_{\lambda_D} v^2 \sin{\beta} \cos{\beta} \lambda}{16 \pi^2 v_s} \ln{\frac{m_D^2}{Q^2}} \nonumber \\
&+& \frac{\lambda_L^4}{48 \pi^2 m_L^4}(-2 A_{\lambda_L}^4 v_s^2 + 24 A_{\lambda_L}^2 m_L^2 v_s^2 - 15 A_{\lambda_L}^3 vs^3 \kappa + 90 A_{\lambda_L} m_L^2 v_s^3 \kappa - 
 36 A_{\lambda_L}^2 v_s^4 \kappa^2 \nonumber \\
&+& 72 m_L^2 v_s^5 \kappa^2 - 35 A_{\lambda_L} v_s^5 \kappa^3 - 12 v_s^6 \kappa^4 + 3 A_{\lambda_L}^3 v_d v_s v_u \lambda - 18 A_{\lambda_L} m_L^2 v_d v_s v_u \lambda \nonumber \\
&+& 24 A_{\lambda_L}^2 v_d v_s^2 v_u \kappa \lambda - 48 m_L^2 v_d v_s^2 v_u \kappa \lambda + 45 A_{\lambda_L} v_d v_s^3 v_u \kappa^2 \lambda + 24 v_d v_s^4 v_u \kappa^3 \lambda \nonumber \\
&-& 9 A_{\lambda_L} v_d^2 v_s v_u^2 \kappa \lambda^2 - 12 v_d^2 v_s^2 v_u^2 \kappa^2 \lambda^2 - A_{\lambda_L} v_d^3 v_u^3 \lambda^3 + 24 m_L^4 v_s^2 \ln{\frac{m_L^2}{\lambda^2 v_s^2}}) + \label{Singlet_Mass} \\
&+& \frac{\lambda_D^4}{32 \pi^2 m_D^4}(-2 A_{\lambda_D}^4 v_s^2 + 24 A_{\lambda_D}^2 m_D^2 v_s^2 - 15 A_{\lambda_D}^3 vs^3 \kappa + 90 A_{\lambda_D} m_D^2 v_s^3 \kappa - 
 36 A_{\lambda_D}^2 v_s^4 \kappa^2 \nonumber \\
&+& 72 m_D^2 v_s^5 \kappa^2 - 35 A_{\lambda_D} v_s^5 \kappa^3 - 12 v_s^6 \kappa^4 + 3 A_{\lambda_D}^3 v_d v_s v_u \lambda - 18 A_{\lambda_D} m_D^2 v_d v_s v_u \lambda \nonumber \\
&+& 24 A_{\lambda_D}^2 v_d v_s^2 v_u \kappa \lambda - 48 m_D^2 v_d v_s^2 v_u \kappa \lambda + 45 A_{\lambda_D} v_d v_s^3 v_u \kappa^2 \lambda + 24 v_d v_s^4 v_u \kappa^3 \lambda \nonumber \\
&-& 9 A_{\lambda_D} v_d^2 v_s v_u^2 \kappa \lambda^2 - 12 v_d^2 v_s^2 v_u^2 \kappa^2 \lambda^2 - A_{\lambda_D} v_d^3 v_u^3 \lambda^3 + 24 m_D^4 v_s^2 \ln{\frac{m_D^2}{\lambda^2 v_s^2}}). \nonumber
\end{eqnarray}
The two leading terms are similar to (\ref{Higgs_Mass_55}), which result from the $\lambda_{eff}$ defined in the previous text, while the $\lambda_{L,D}^4$ terms, especially the $24 m_{L,D}^4 v_s^2 \ln{\frac{m_{L,D}^2}{\lambda^2 v_s}}$ reflect the fact that the mass of the singlet Higgs also receive the corrections from the mass hierarchy of the corresponding vector-like fermions and sfermions, which is Q-independent.

\begin{figure}
\includegraphics[width=4in]{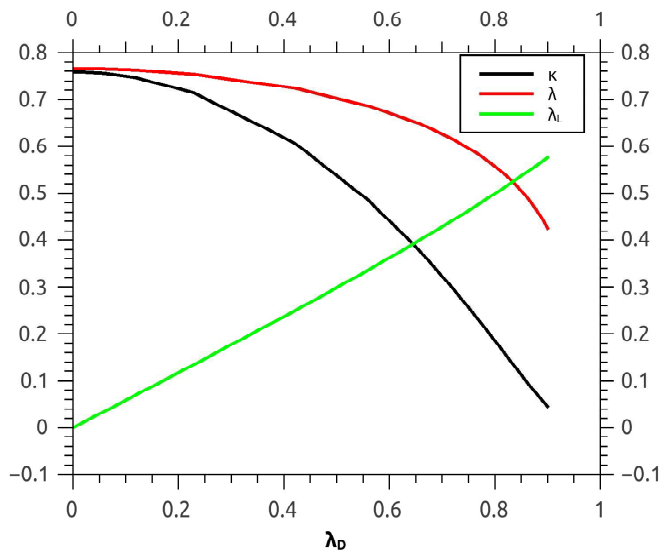}
\caption{The coupling constants near Q=1TeV in the boundary condition that $\lambda(Q_{GUT})=\kappa(Q_{GUT})=3$ and $\lambda_D(Q_{GUT})=\lambda_L(Q_{GUT})$. The GUT-scale is defined as $Q_{GUT}=1.81 \times 10^{16} \text{GeV}$, and different values of $\lambda_D(Q_{GUT})=\lambda_L(Q_{GUT})$ are taken into the RG input parameters.}
\label{lDlLlk}
\end{figure}

To see the possible mass-spectrum of the vector-like fermions, we look into the RG trajectories of the coupling constants. We can learn from (\ref{Gauge_RG}) that gauge terms contribute negative values to all yukawa RG-$\beta$ functions, while the yukawa terms always contribute positive ones. $S$ is a SM gauge-singlet, so the lack of minus terms decides the quasi-fixing point of $\kappa$ to be actually 0. $H_u$ and $H_d$ are not SM gauge-singlets, however, we coupled a lot of things on $S$ and if $\lambda_D$ and $\lambda_L$ are too large, $\lambda$ also tend to be small. At the GUT point, if we set $\lambda(Q_{GUT})=\kappa(Q_{GUT})=3$, which is near the perturbative limit of $\sqrt{4 \pi}$, and apply $\lambda_D(Q_{GUT})=\lambda_L(Q_{GUT})$ with different values, and then we run the RG-trajectories down, we can see the relationship between these coupling constants near 1TeV through Figure \ref{lDlLlk}.

\begin{figure}
\includegraphics[width=3in]{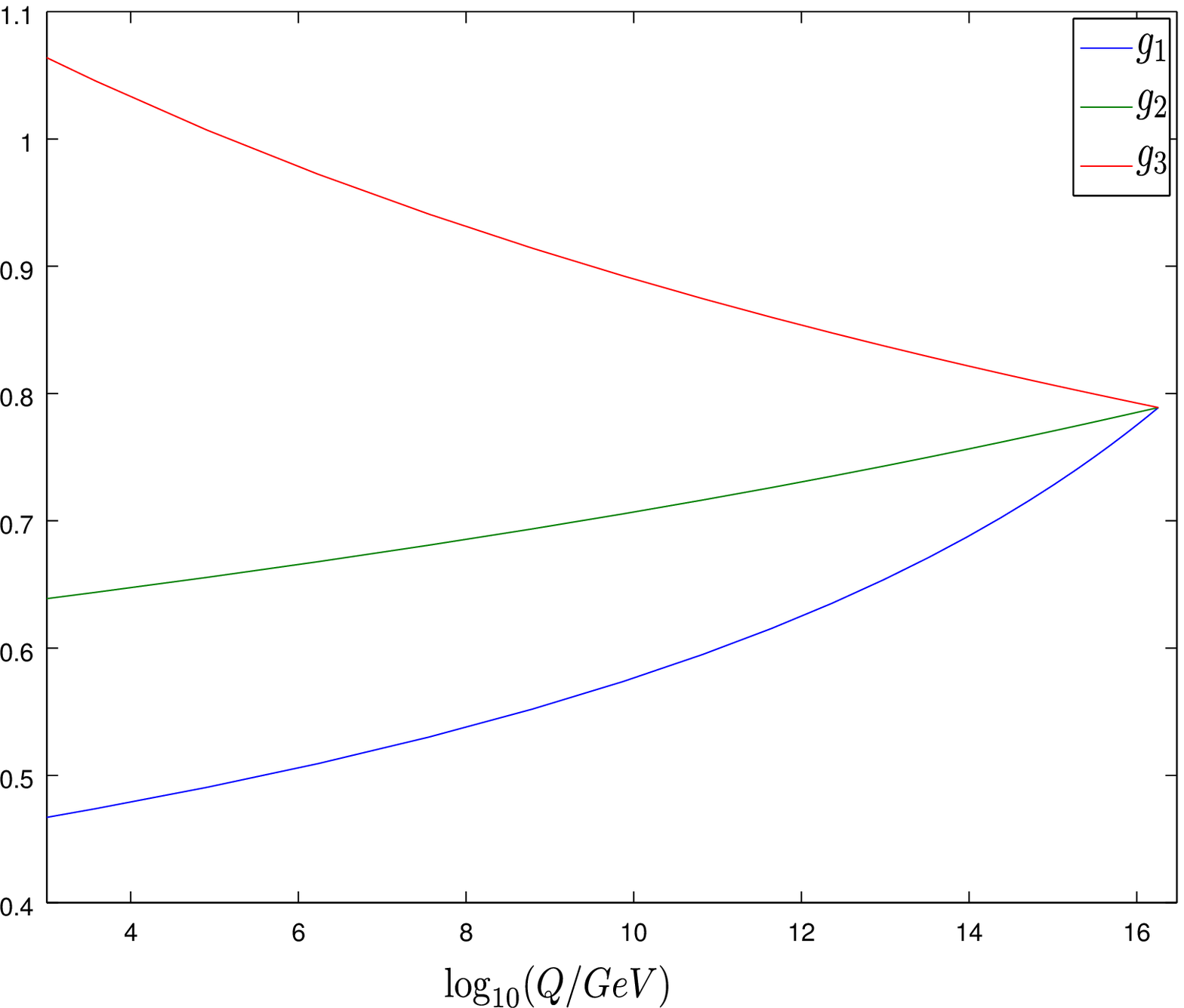}
\includegraphics[width=3in]{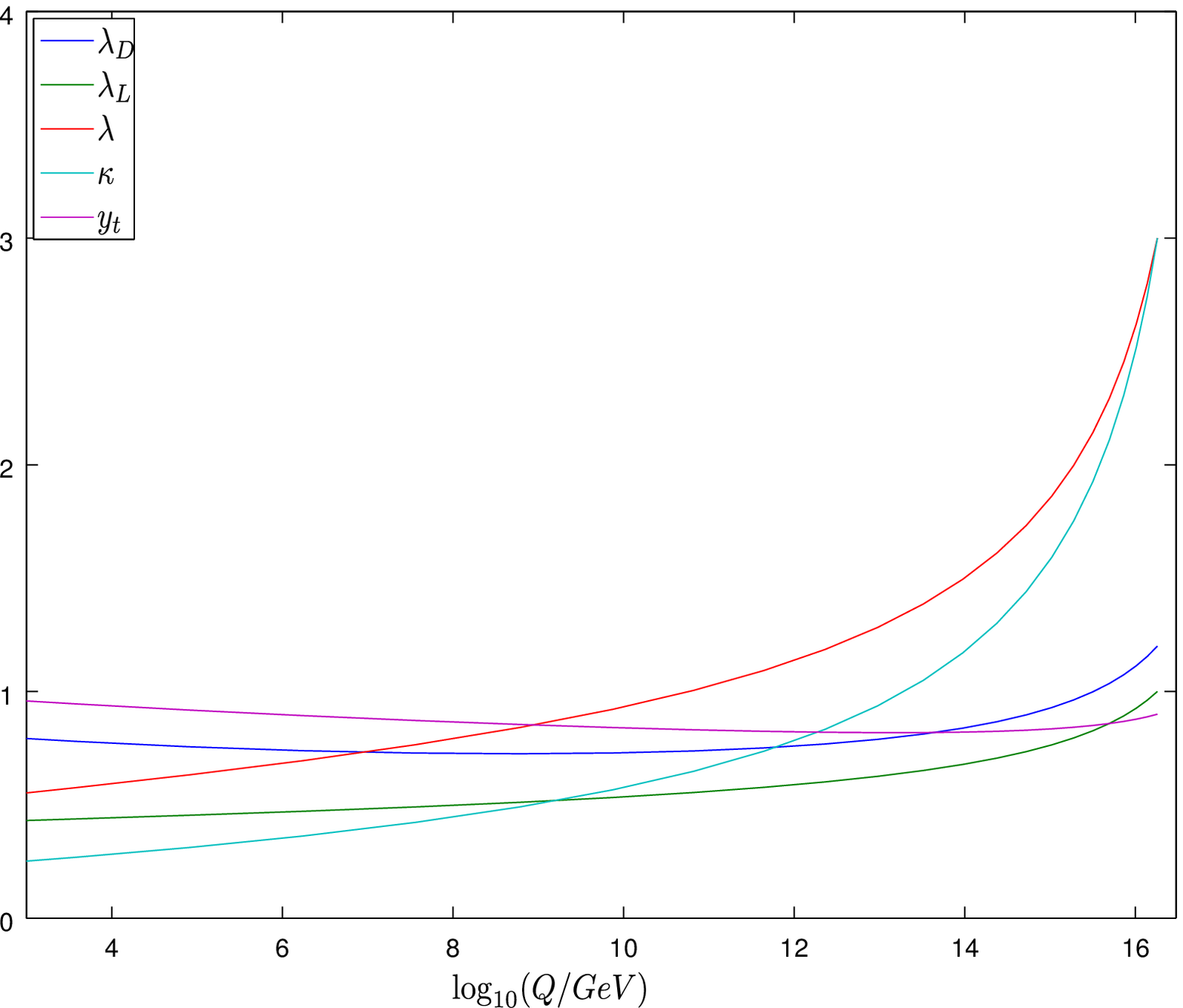}
\caption{This figure running couplings in $5+\overline{5}$ theory. The left panel shows the trajectories of the gauge couplings. Notice that they are converged into one point accurately. The right panel shows the corresponding yukawa couplings. It is the yukawa couplings that contribute into the gauge RG-$\beta$ functions so that they can converge into one point at the scale $Q=1.81 \times 10^{16} \text{GeV}$.}
\label{Trajectory_55bar}
\end{figure}

From the two-loop $\beta$ functions of gauge couplings listed in (\ref{Gauge_RG}), we can learn that the yukawa coupling constants play crucial roles if ever they're large enough. It is ever-known that the unification of gauge coupling constants is not that good even in the circumstance of supersymmetry, although it has been improved a lot when compared with the case of SM. If we want to adjust the yukawa couplings in order to drive the gauge couplings into unifying in MSSM or NMSSM, there does not left much room in the parameter space because we do not have many notable yukawa coupling constants to be adjusted, and the top yukawa $Q_3 H_u t_3$ affects on all $g_1$, $g_2$, $g_3$, making it difficult to converge the intersection points. In our case, $\lambda_D$ strongly influences the trajectory of $g_3$ however slightly modifies $g_1$ due to $D$ and $\bar{D}$'s relatively small super-charge $\frac{1}{3}$, and $\lambda_L$ only effects on $g_1$ and $g_2$, so we can move the intersection point separately by adjusting $\lambda_D$ and $\lambda_L$. After several attempts, we can reach a boundary condition
\begin{eqnarray}
&&g_1=g_2=g_3=0.789,\text{\quad}\lambda_D=1.2,\text{\quad}\lambda_L=1,\text{\quad}\lambda=\kappa=3,\text{\quad}y_t=0.9, \nonumber \\
&&\text{ at the scale }Q=1.81\times 10^{16} \text{GeV},
\end{eqnarray}
and if we run down into Q=1TeV, the gauge coupling constants are accurately in accordance with the low-energy data $g_1(1\text{TeV})=0.4670$, $g_2(1\text{TeV})=0.6388$, $g_3(1\text{TeV})=1.063$. See Figure \ref{Trajectory_55bar} for the trajectories.

\section{$10+\overline{10}$ Model}

Without the help of extra "vector-like neutrino" N, the $5+\overline{5}$ model can only contribute to the SM-like higgs mass through S. However, the $10+\overline{10}$ model contains direct vertices $Q H_u U$ and $\bar{Q} H_d \bar{U}$. Unlike the $5+\overline{5}$ case, the $10+\overline{10}$ model contains four $\frac{2}{3}$ charged squarks and thus a $4 \times 4$ matrix needs to be diagonalized, so it a simple analytical solution does not exist.

The $4 \times 4$ mass-squared matrix of up-type squarks is shown below,
\begin{eqnarray}
&&\left[
\begin{array}{cccc}
\lambda_Q^2 v_s^2 + y_u^2 v_u^2 + m_Q^2 & \kappa \lambda_Q v_s^2 - \lambda \lambda_Q v_u v_d & -y_u \lambda v_d v_s & -y_{\bar{u}} \lambda_Q v_d v_s + \lambda_U y_{\bar{u}} v_u v_s \\
\kappa \lambda_Q v_s^2 - \lambda \lambda_Q v_u v_d & \lambda_Q^2 v_s^2 + y_d^2 v_d^2 + m_Q^2 & -\lambda_U y_{\bar{u}} v_d v_s + \lambda_Q y_u v_u v_s & \lambda y_{\bar{u}} v_u v_s \\
-y_u \lambda v_d v_s & -\lambda_U y_{\bar{u}} v_d v_s + \lambda_Q y_u v_u v_s & \lambda_U^2 v_s^2 + y_u^2 v_u^2 + m_Q^2 & \kappa \lambda_U v_s^2 - \lambda \lambda_U v_u v_d \\
-y_{\bar{u}} \lambda_Q v_d v_s + \lambda_U y_{\bar{u}} v_u v_s & \lambda y_{\bar{u}} v_u v_s & \kappa \lambda_U v_s^2 - \lambda \lambda_U v_u v_d & \lambda_U^2 v_s^2 + v_d^2 y_{\bar{u}}^2 + m_Q^2
\end{array}
\right]\nonumber \\
&+&\left[
\begin{array}{cccc}
0 & A_{\lambda_Q} \lambda_Q v_s & A_{y_u} y_u v_u & 0 \\
A_{\lambda_Q} \lambda_Q v_s & 0 & 0 & -A_{y_{\bar{u}}} y_{\bar{u}} v_d \\
A_{y_u} y_u v_u & 0 & 0 & A_{\lambda_U} \lambda_U v_s \\
0 & -A_{y_{\bar{u}}} y_{\bar{u}} v_d & A_{\lambda_U} \lambda_U v_s & 0
\end{array}
\right], \label{SU_Matrix_1010bar}
\end{eqnarray}
where we can observe that unlike the consequence of MSSM+($10+\overline{10}$)VL model (e.g. in \cite{Martin:2010dc}), lots of off-diagonal terms automatically appear, so the diagonalizing process becomes much more difficult. The mass-squared matrix of the two down-type squarks is
\begin{eqnarray}
\left[
\begin{array}{cc}
  m_Q^2+\lambda_Q^2 v_s^2 & - v_s^2 \kappa \lambda_Q + v_d v_u \lambda \lambda_Q - A_{\lambda_Q} \lambda_Q v_s \\
  -v_s^2 \kappa \lambda_Q + v_u v_d \lambda \lambda_Q - A_{\lambda_Q} \lambda_Q v_s & m_Q^2 + \lambda_Q^2 v_s^2
\end{array}
\right]. \label{SD_Matrix_1010bar}
\end{eqnarray}
The mass matrix of two VL fermionic up-type quark is
\begin{eqnarray}
\left[
\begin{array}{cc}
\lambda_Q v_s & y_u v_u \\
-y_{\bar{u}} v_d & \lambda_U v_s
\end{array}
\right], \label{FU_Matrix_1010bar}
\end{eqnarray}
while there is only one VL down-type quark
\begin{eqnarray}
M_{Q_D}=\lambda_Q v_s. \label{FD_Matrix_1010bar}
\end{eqnarray}

Direct calculation diagonalizing (\ref{SU_Matrix_1010bar}) is so lengthy and troublesome, so we expand the result in series of $y_u$, $y_{\bar{u}}$, $\lambda$ and $\kappa$, and set $A_{y_u}=A_{y_{\bar{u}}}$, $A_{\lambda_Q}=A_{\lambda_U}$. According to experience, the coupling constants of leptons is usually smaller than quarks because leptons do not have colors, thus their quasi-fixing points are smaller. The leptons also do not receive $N_C$ accumulations, so, for simplicity, we ignore all leptonic contributions here. If we would like a relatively large $\tan\beta$, say, $\tan\beta>2$, the SM-like lightest higgs will mainly be $H_u^0$ and thus $y_{\bar{u}}$ can also be ignored. Let's define
\begin{eqnarray}
\delta \lambda_Q=\lambda_U-\lambda_Q,
\end{eqnarray}
and expand the final result according to $y_u$, $\delta \lambda_Q$, $\lambda$, $\kappa$. Similar to the process in (\ref{Higgs_Mass_55}), we acquire
\begin{eqnarray}
\Delta m_h^2 &=& \frac{1}{16 \pi^2 M_{\tilde{Q}}^4} v^2 \sin^2 \beta \left[ 36 M_{FQ}^4 \lambda^2 \lambda_Q^2 \cos^2 \beta \ln\left( \frac{M_{\tilde{Q}}^2}{Q^2} \right) \right. \nonumber \\
&-&  y_u^2 \sin^2 \beta \left( A_{y_u}^4 + 2 A_{y_u}^2 M_{FQ}^2 + M_{FQ}^4 - 12 A_{y_u}^2 M_{\tilde{Q}}^2 - 6 M_{FQ}^2 M_{\tilde{Q}}^2 + 10 M_{\tilde{Q}}^4 \right) \nonumber \\
&+& \left. 12 y_u^2 \sin^2 \beta M_{\tilde{Q}}^2 \ln \left( \frac{M_{\tilde{Q}}^2}{M_{FQ}^2} \right) - 6 y_u^2 A_{y_u} M_{FQ} M_{\tilde{Q}}^2 \lambda \lambda_Q \sin (2 \beta) \right], \label{mHiggs_1010bar}
\end{eqnarray}
where $M_{FQ}=\lambda_Q v_s$, $M_{\tilde{Q}}=\sqrt{m_Q^2+\lambda_Q^2 v_s^2}$ are the estimated masses of fermionic and bosonic up-type quarks. We cut the series up to $y_u^4$, $\delta \lambda^2$, $\lambda^2$ and $\kappa^2$. However, $\delta \lambda$ disappears in the final result, telling us that the difference between $\lambda_Q$ and $\lambda_U$ does not exert a large effect on SM-like higgs mass.

Similar to (\ref{Singlet_Mass}), the singlet Higgs also receives one-loop quantum corrections. However, in spite of the similar $\lambda_Q^2$, $\lambda_U^2$ terms, the Q-independent terms are so complicated, that we don't show them in this paper.

\begin{figure}
\includegraphics[width=3in]{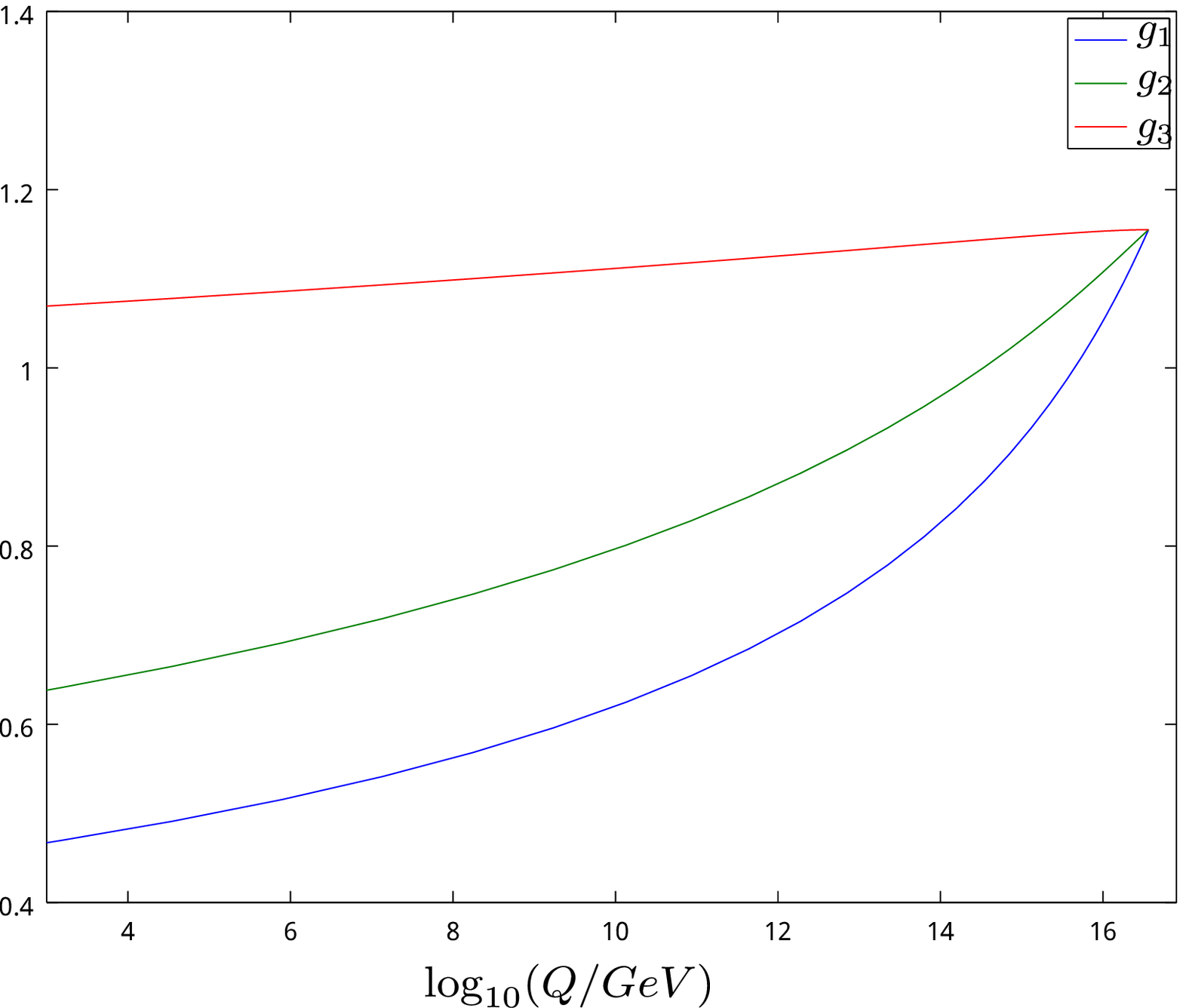}
\includegraphics[width=3in]{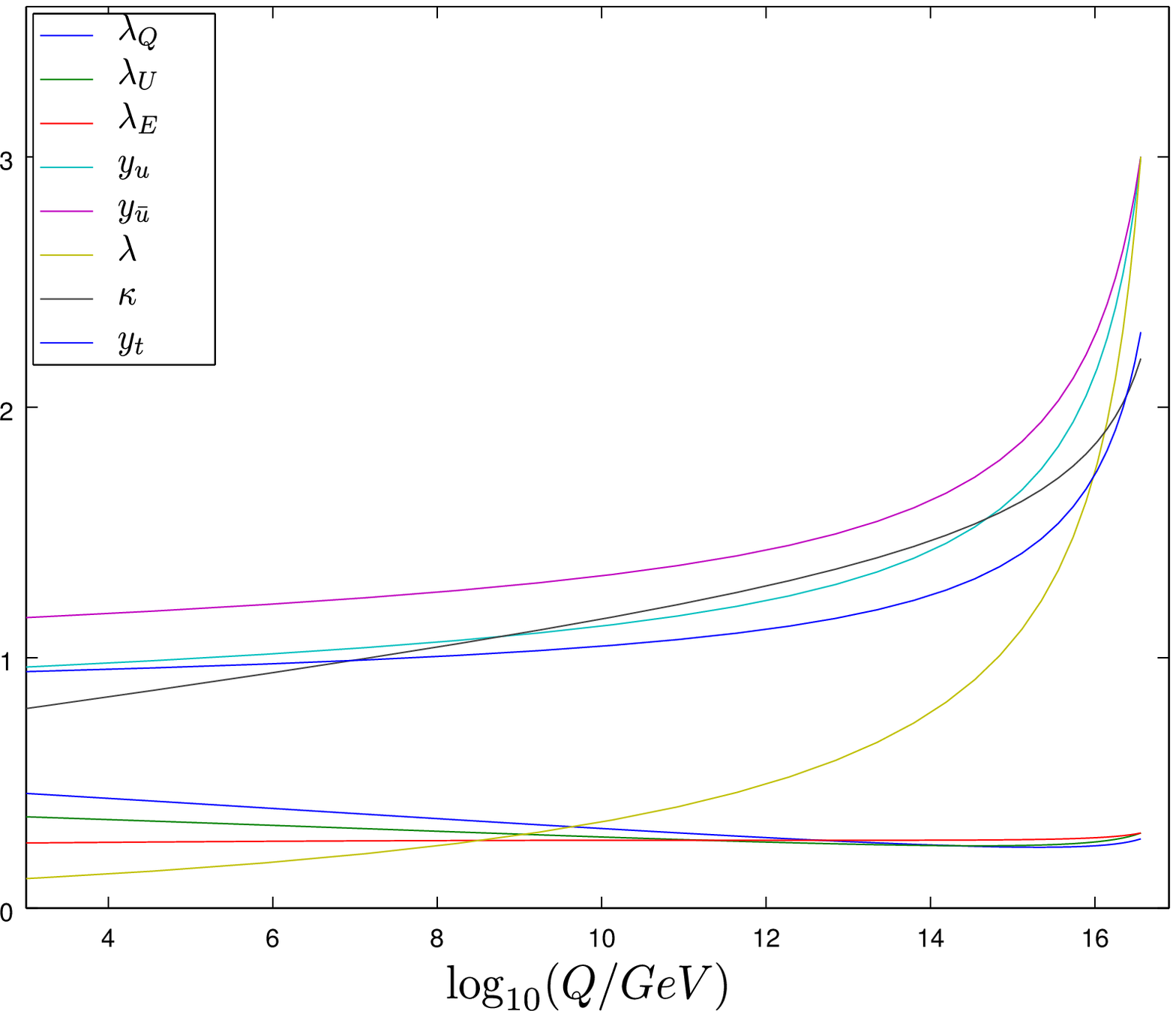}
\caption{The coupling constant trajectories of $10+\overline{10}$ theory. The left panel shows the gauge couplings, and the right panel shows the yukawa couplings.}
\label{RG_1010bar}
\end{figure}
Now we are going to unify the gauge couplings. It is much more difficult to converge the intersection points in this circumstance than $5+\overline{5}$ model, because the gauge couplings run into a larger value, $\sim 1.2$, which is much less sensitive to the adjusting of the large yukawa couplings. If we set
\begin{eqnarray}
&&g_1=g_2=g_3=1.155,\text{\quad}\lambda_Q=0.27657,\text{\quad}\lambda_U=0.3,\text{\quad}\lambda_E=0.3,\text{\quad} \nonumber \\
&&y_u=y_{\bar{u}}=3,\text{\qquad}\lambda=3,\text{\quad}\kappa=2.19446,\text{\quad}y_t=2.3,\nonumber \\
&&\text{ at scale }Q=3.62 \times 10^{16}\text{GeV},
\end{eqnarray}
after running down to $Q=1000$GeV, we get $g_1=0.4670$, $g_2=0.6382$, $g_3=1.069$. See Fig \ref{RG_1010bar} for trajectories. There is a little deviation from the (\ref{g_1TeV}) in Appendix A.

\section{$5+\overline{5}+10+\overline{10}$ Model}

\begin{figure}
\includegraphics[width=3in]{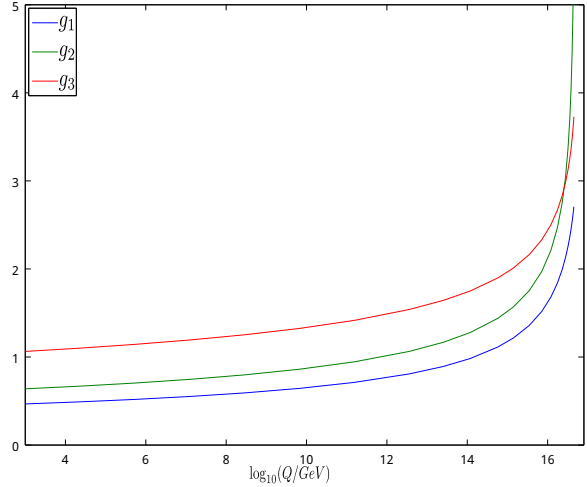}
\caption{The two-loop trajectories of three gauge coupling constants under $5+\overline{5}+10+\overline{10}$+NMSSM theory. All yukawa couplings are closed and the trajectories are all run from $Q=1$TeV}
\label{NoYukawa}
\end{figure}

If we ignore all the yukawa terms and put all the extra particles beyond SM at $1$TeV, and calculate the gauge RG-$\beta$ functions up to 2-loop level, the $g_2$ trajectory actually blows up before $g_1$ can meet $g_3$. See Fig \ref{NoYukawa}.

Then we can add up yukawa couplings to modify the coupling constants' trajectories. Now that $g_2$ runs the fastest, we add up $\lambda_Q$, $\lambda_L$, $y_{u,\bar{u},d,\bar{d},E,\bar{E}}$ to press the $g_2$-$\beta$ function. However, the RG-equations are not stable enough if we run from $Q=1$TeV upwards to GUT-scale. If we apply the relatively large yukawa coupling constants to ``press'' the gauge RG-$\beta$ functions, it is easy for the yukawa coupling constants to blow up before the GUT is reached. However, it is much better to run from GUT-scale downwards to $1$TeV, and by adjusting the yukawa coupling constants, we can acquire the correct values near $Q=1$TeV.

There is another severe problem that $S H_u H_d$ receives so many corrections through the self-energy diagrams on $S$. These corrections depress $\lambda$ very much, forcing the $v_s$ to be extremely large. If we want to discuss the VL particles of roughly $1$TeV, this is not a good news, because this requires $\lambda_{Q,U,D,L,E}$ to be rather small.

\begin{figure}
\includegraphics[width=3in]{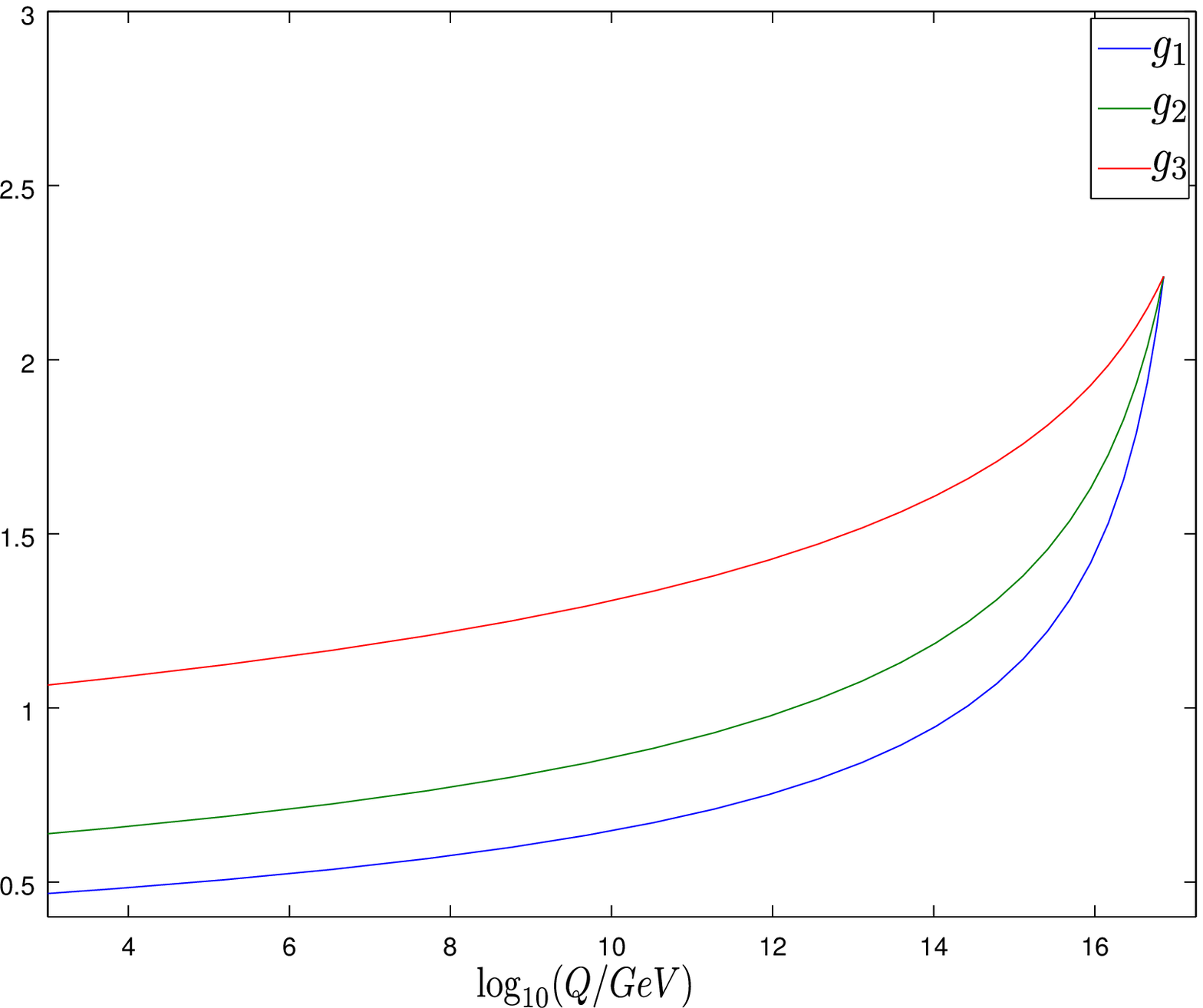}
\includegraphics[width=3in]{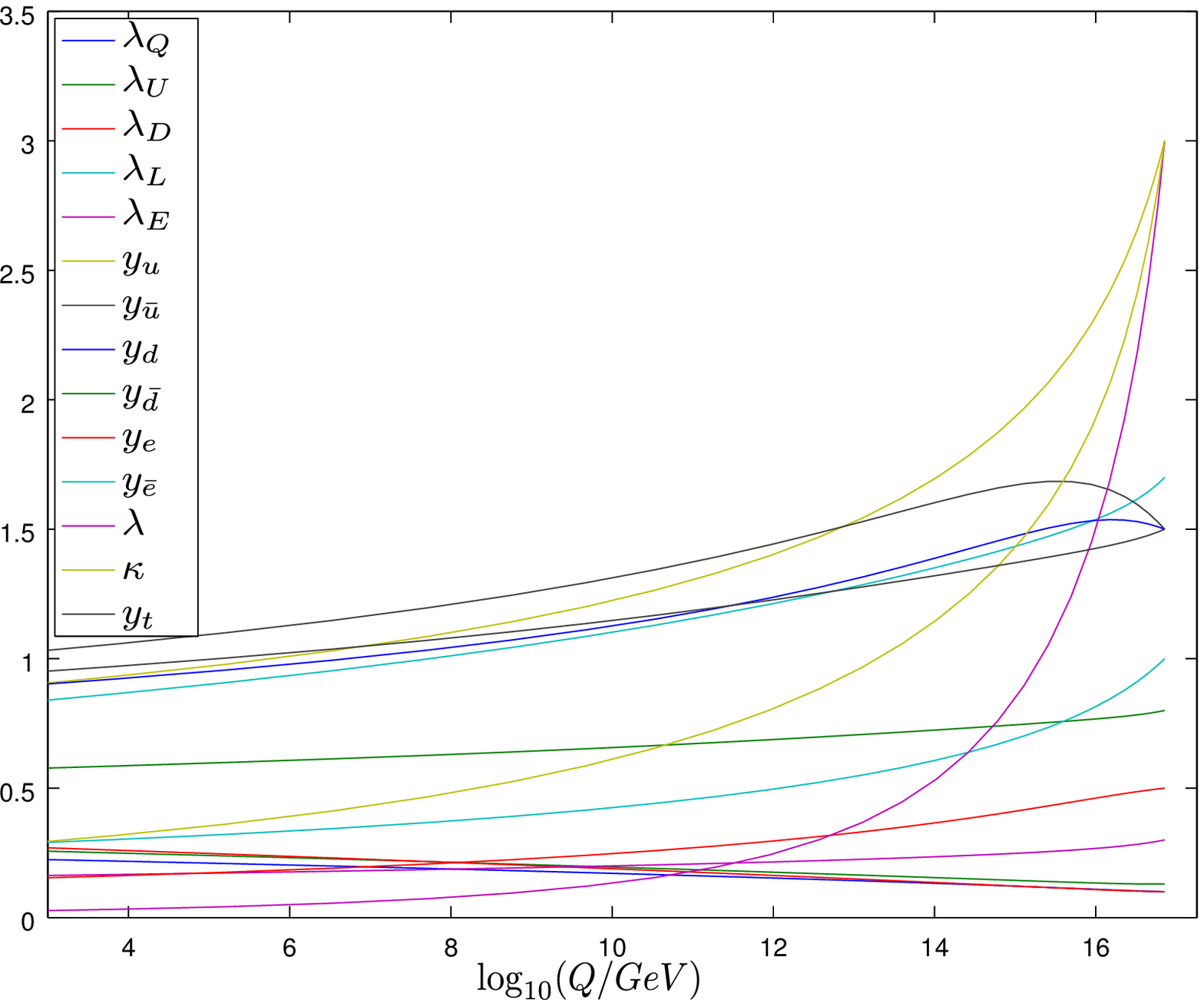}
\caption{The coupling constant trajectories of $5+\overline{5}+10+\overline{10}$ theory. The left panel show the gauge couplings, and the right panel shows the yukawa couplings.}
\label{RG_55bar1010bar}
\end{figure}

However, finally, we are still able to get a group of parameters, with the $\lambda(Q=1\text{TeV}) \sim 0.03$, and $\lambda_Q(Q=1\text{TeV}) \sim 0.2$. If we set $\mu_{eff} ~ 200\text{GeV}$, $M_Q \sim 1.5\text{TeV}$ can still be reached. We are also able to converge the intersection points, and all of the GUT-scale $g_1$, $g_2$, and $g_3$ are smaller than $\sqrt{4 \pi} \sim 3$. See Figure \ref{RG_55bar1010bar}, the boundary conditions are
\begin{eqnarray}
&& g_1=g_2=g_3=2.239,\qquad \lambda_Q=0.1,\text{\qquad}\lambda_U=0.13,\text{\qquad}\lambda_D=0.1,\nonumber \\
&&\lambda_L=1.7,\text{\qquad}\lambda_E=0.3,\text{\qquad}y_u=3,\text{\qquad}y_{\bar{u}}=1.5,\text{\qquad}y_d=1.5,\text{\qquad}y_{\bar{d}}=0.8,\nonumber \\
&&y_e=0.5,\text{\qquad}y_{\bar{e}}=1,\text{\qquad}\lambda=3,\text{\qquad}\kappa=3,\text{\qquad}y_t=1.5, \nonumber \\
&&\text{at the scale } Q=7.15 \times 10^{16} \text{GeV}
\end{eqnarray}

The mass-squared matrices of the up-type squarks and the down-type squarks are both $4 \times 4$ shaped, so the complete formulae of $\Delta m_{h}^2$ and $\Delta m_{S}^2$ are too complicated to be shown in this paper. For SM-like Higgs, naively, we believe that there are two vertices involving $H_u$: the $Q H_u U$ and $\bar{Q} H_u \bar{D}$, so the contribution similar to the (\ref{mHiggs_1010bar}) can be doubled. However, if we enlarge both $y_u$ and $y_{\bar{d}}$, the $\lambda$ would be so small that we cannot get TeV scale VL-quarks. In order to accumulate the Higgs mass, we could only choose one of them to be large while giving up another.

\section{Conclusion}

By combining supersymmetric vector-like theory and NMSSM, and couple the $S$ with the vector-like particles, we can give a natural source of the vector-like mass after $Z_3$ breaks. The mass of the VL fermions is of the similar quantity as the vev of $\tilde{S}$ in the range of $1\sim 100$TeV. The new yukawa couplings invented in this theory can help unifying the gauge-coupling constants due to their contribution to the two-loop gauge RG-$\beta$ functions. It is usually difficult to reach GUT before Landau pole in a TeV-scale MSSM+$5+\overline{5}+10+\overline{10}$ model. However, with the help of the yukawa couplings appeared in our models, we succeeded in converging the trajectories of the gauge coupling constants before they blow up. The coupling between VL particles and higgs can also contribute into the higgs mass. Unlike MSSM+VL models, in our model, even $5+\overline{5}$ influences the effective higgs mass through the couplings between VL particles and the S. We have calculated the contributions to higgs masses analytically in NMSSM+$5+\overline{5}$ model, and only to SM-like higgs mass in NMSSM+$10+\overline{10}$ model, and have discussed it briefly in NMSSM+$5+\overline{5}+10+\overline{10}$ model.

\begin{acknowledgements}

We would like to thank Professor Chun Liu, Dr$.$ Jia-shu Lu, Mr$.$ Weicong Huang for helpful discussions.
This work was supported in part by the National Natural Science
Foundation of China under Nos. 11375248, and by the
National Basic Research Program of China under Grant No. 2010CB833000.

\end{acknowledgements}

\appendix
\section{The RG-$\beta$ functions}

Under $1$TeV, we run the gauge coupling constants through one-loop SM functions \cite{SUSY_GUT}
\begin{eqnarray}
\alpha_1^{-1} (Q) &=& \alpha_1^{-1} (m_Z) - \frac{41}{20 \pi} \ln \frac{Q}{m_Z} \nonumber \\
\alpha_2^{-1} (Q) &=& \alpha_2^{-1} (m_Z) + \frac{19}{12 \pi} \ln \frac{Q}{m_Z} \\
\alpha_3^{-1} (Q) &=& \alpha_3^{-1} (m_Z) + \frac{7}{2 \pi} \ln \frac{Q}{m_Z}. \nonumber
\end{eqnarray}
We start from
\begin{eqnarray}
\alpha_1(m_Z) &=& 0.0169 \nonumber \\
\alpha_2(m_Z) &=& 0.0338 \\
\alpha_3(m_Z) &=& 0.1184, \nonumber
\end{eqnarray}
where $m_Z=91.2$GeV, and all the data is calculated according to the values in \cite{PDG} and then get
\begin{eqnarray}
g_1(Q=1\text{TeV})=0.4670 \nonumber \\
g_2(Q=1\text{TeV})=0.6388 \label{g_1TeV}\\
g_3(Q=1\text{TeV})=1.0633. \nonumber
\end{eqnarray}

Upon $1$TeV, we calculate the $\beta$ functions according to the steps listed in \cite{SUSYRun_1, SUSYRUN_2}. The gauge RG-$\beta$ functions are calculated up to two-loop accuracy and the yukawa RG-$\beta$ functions are calculated up to one-loop. 

The gauge RG-$\beta$ are listed below
\begin{eqnarray}
\beta_{g_1} &=& \frac{g_1^3}{16 \pi^2} \left( \frac{33}{5} + \frac{N_E}{2} \right) + \frac{g_1^3}{(16 \pi^2)^2} \left[ \sum_j \left(BG_{1j} g_j^2\right) - C_{\lambda_Q 1} \lambda_Q - C_{\lambda_U 1} \lambda_U - C_{\lambda_D 1} \lambda_D \right. \nonumber \\ 
&-& \left. C_{\lambda_L 1} \lambda_L - C_{\lambda_E 1} \lambda_E - C_{y_u 1} y_u - C_{y_{\bar{u}} 1} y_{\bar{u}} - C_{y_d 1} y_d - C_{y_{\bar{d}} 1} y_{\bar{d}} - C_{y_e 1} y_e - C_{y_{\bar{e}} 1} y_e \right. \nonumber \\
&-& \left. C_{\lambda 1} \lambda - C_{y_t 1} y_t \right] \nonumber \\
\beta_{g_2} &=& \frac{g_2^3}{16 \pi^2} \left( 1 + \frac{N_E}{2} \right) + \frac{g_2^3}{(16 \pi^2)^2} \left[ \sum_j \left(BG_{2j} g_j^2\right) - C_{\lambda_Q 2} \lambda_Q - C_{\lambda_U 2} \lambda_U - C_{\lambda_D 2} \lambda_D \right. \nonumber \\ 
&-& \left. C_{\lambda_L 2} \lambda_L - C_{\lambda_E 2} \lambda_E - C_{y_u 2} y_u - C_{y_{\bar{u}} 2} y_{\bar{u}} - C_{y_d 2} y_d - C_{y_{\bar{d}} 2} y_{\bar{d}} - C_{y_e 2} y_e - C_{y_{\bar{e}} 2} y_{\bar{e}} \right. \nonumber \\
&-& \left. C_{\lambda 2} \lambda - C_{y_t 2} y_t \right] \label{Gauge_RG}\\
\beta_{g_3} &=& \frac{g_3^3}{16 \pi^2} \left( -3 + \frac{N_E}{2} \right) + \frac{g_3^3}{(16 \pi^2)^2} \left[ \sum_j \left(BG_{3j} g_j^2\right) - C_{\lambda_Q 3} \lambda_Q - C_{\lambda_U 3} \lambda_U - C_{\lambda_D 3} \lambda_D \right. \nonumber \\ 
&-& \left. C_{\lambda_L 3} \lambda_L - C_{\lambda_E 3} \lambda_E - C_{y_u 3} y_u - C_{y_{\bar{u}} 3} y_{\bar{u}} - C_{y_d 3} y_d - C_{y_{\bar{d}} 3} y_{\bar{d}} - C_{y_e 3} y_e - C_{y_{\bar{e}} 3} y_{\bar{e}} \right. \nonumber \\
&-& \left. C_{\lambda 3} \lambda - C_{y_t 3} y_t \right] \nonumber
\end{eqnarray}
where
\begin{eqnarray}
C_{\lambda_Q}=[2/5,6,4] \nonumber \\
C_{\lambda_U}=[16/5,0,2] \nonumber \\
C_{\lambda_D}=[4/5,0,2] \nonumber \\
C_{\lambda_L}=[6/5,2,0] \nonumber \\
C_{\lambda_E}=[12/5,0,0] \nonumber \\
C_{y_u}=[26/5,6,4] \nonumber \\
C_{y_{\bar{u}}}=[26/5,6,4] \nonumber \\
C_{y_d}=[14/5,6,4] \nonumber \\
C_{y_{\bar{d}}}=[14/5,6,4] \nonumber \\
C_{y_e}=[18/5,2,0] \nonumber \\
C_{y_{\bar{e}}}=[18/5,2,0] \nonumber \\
C_{\lambda}=[6/5,2,0] \nonumber \\
C_{y_t}=[26/5,6,4] \nonumber
\end{eqnarray}
and
\begin{eqnarray}
\left[ BG^{MSSM}_{ij} \right]&=&\left[
\begin{array}{ccc}
199/25 & 27/5 & 88/5 \\
9/5 & 25 & 24 \\
11/5 & 9 & 14
\end{array} \right] \nonumber \\
\left[ BG^5_{ij} \right] &=& \left[
\begin{array}{ccc}
7/30 & 9/10 & 16/15 \\
3/10 & 7/2 & 0 \\
2/15 & 0 & 17/3
\end{array} \right] \nonumber \\
\left[ BG^{10}_{ij} \right] &=& \left[
\begin{array}{ccc}
23/10 & 3/10 & 24/5 \\
1/10 & 21/2 & 8 \\
3/5 & 3 & 17
\end{array} \right], \nonumber
\end{eqnarray}
so
\begin{eqnarray}
BG_{ij}&=&BG^{MSSM}_{ij}+2 BG^5_{ij} \text{, for $5+\overline{5}$ theory,} \nonumber \\
BG_{ij}&=&BG^{MSSM}_{ij}+2 BG^{10}_{ij} \text{, for $10+\overline{10}$ theory,} \nonumber \\
BG_{ij}&=&BG^{MSSM}_{ij}+2 BG^5_{ij}+2 BG^{10}_{ij} \text{, for $5+\overline{5}+10+\overline{10}$ theory.} \nonumber
\end{eqnarray}

The RG-$\beta$ functions of the yukawa couplings are
\begin{eqnarray}
\beta_{\lambda_Q}&=&\frac{1}{16 \pi^2} \lambda_Q [ 8 \lambda_Q^2 + 3 \lambda_U^2 + 3 \lambda_D^2 + 2 \lambda_L^2 + \lambda_E^2 + y_{\bar{u}}^2 + y_{\bar{d}}^2 + y_u^2 + y_d^2 + 2 \lambda^2 \nonumber \\
&-&	 2 (\frac{1}{60} g_1^2 + \frac{3}{4} g_2^2 + \frac{4}{3} g_3^2)] \nonumber
\end{eqnarray}
\begin{eqnarray}
\beta_{\lambda_U}&=&\frac{1}{16 \pi^2} \lambda_U [ 5 \lambda_U^2 + 6 \lambda_Q^2 + 3 \lambda_D^2 + 2 \lambda_L^2 + \lambda_E^2 + y_{\bar{u}}^2 + y_u^2 + 2 \lambda^2 \nonumber \\
&-&	 2 (\frac{4}{15} g_1^2 + \frac{4}{3} g_3^2)] \nonumber
\end{eqnarray}
\begin{eqnarray}
\beta_{\lambda_D}&=&\frac{1}{16 \pi^2} \lambda_D [ 5 \lambda_D^2 + 6 \lambda_Q^2 + 3 \lambda_U^2 + 2 \lambda_L^2 + \lambda_E^2 + y_{\bar{d}}^2 + y_d^2 + 2 \lambda^2 \nonumber \\
&-&	 2 (\frac{1}{15} g_1^2 + \frac{4}{3} g_3^2)] \nonumber
\end{eqnarray}
\begin{eqnarray}
\beta_{\lambda_L}&=&\frac{1}{16 \pi^2} \lambda_L [ 4 \lambda_L^2 + 6 \lambda_Q^2 + 3 \lambda_D^2  + \lambda_E^2 + y_{\bar{e}}^2 + y_e^2 + 2 \lambda^2 \nonumber \\
&-&	 2 (\frac{3}{20} g_1^2 + \frac{3}{4} g_2^2)] \nonumber
\end{eqnarray}
\begin{eqnarray}
\beta_{\lambda_E}&=&\frac{1}{16 \pi^2} \lambda_E [ 3 \lambda_E^2 + 3 \lambda_Q^2 + 3 \lambda_U^2 + 3 \lambda_D^2 + 2 \lambda_L^2 + 2 y_e^2 + 2 y_{\bar{e}}^2 + 2 \lambda^2 \nonumber \\
&-&	 2 (\frac{6}{5} g_1^2)] \nonumber
\end{eqnarray}
\begin{eqnarray}
\beta_{y_u}&=&\frac{1}{16 \pi^2} y_u [ 6 y_u^2 + 3 y_{\bar{d}}^2 + y_d^2 + \lambda_U^2 +  \lambda_Q^2 + y_{\bar{e}}^2 + \lambda^2 + 3 y_t^2 \nonumber \\
&-&	 2 (\frac{13}{30} g_1^2 + \frac{3}{2} g_2^2 + \frac{8}{3} g_3^2)] \nonumber
\end{eqnarray}
\begin{eqnarray}
\beta_{y_{\bar{u}}}&=&\frac{1}{16 \pi^2} y_{\bar{u}} [ 6 y_{\bar{u}}^2 + y_{\bar{d}}^2 + 3 y_d^2 + \lambda_U^2 +  \lambda_Q^2 + y_e^2 + \lambda^2 \nonumber \\
&-&	 2 (\frac{13}{30} g_1^2 + \frac{3}{2} g_2^2 + \frac{8}{3} g_3^2)] \nonumber
\end{eqnarray}
\begin{eqnarray}
\beta_{y_d}&=&\frac{1}{16 \pi^2} y_d [ 6 y_d^2 + y_u^2 + 3 y_{\bar{u}}^2 + y_e^2 + \lambda_D^2 +  \lambda_Q^2 + \lambda^2  \nonumber \\
&-&	 2 (\frac{7}{30} g_1^2 + \frac{3}{2} g_2^2 + \frac{8}{3} g_3^2)] \nonumber
\end{eqnarray}
\begin{eqnarray}
\beta_{y_{\bar{d}}}&=&\frac{1}{16 \pi^2} y_{\bar{d}} [ 6 y_{\bar{d}}^2 + 3 y_u^2 + y_{\bar{u}}^2 + y_{\bar{e}^2} + \lambda_D^2 +  \lambda_Q^2 + \lambda^2 + 3 y_t^2 \nonumber \\
&-&	 2 (\frac{7}{30} g_1^2 + \frac{3}{2} g_2^2 + \frac{8}{3} g_3^2)] \nonumber
\end{eqnarray}
\begin{eqnarray}
\beta_{y_e}&=&\frac{1}{16 \pi^2} y_e [ 4 y_e^2 + 3 y_{\bar{u}}^2 + 3 y_d^2 + \lambda_L^2 +  \lambda_E^2 + \lambda^2 + 3 y_t^2 \nonumber \\
&-&	 2 (\frac{9}{10} g_1^2 + \frac{3}{2} g_2^2 )] \nonumber
\end{eqnarray}
\begin{eqnarray}
\beta_{y_{\bar{e}}}&=&\frac{1}{16 \pi^2} y_{\bar{e}} [ 4 y_{\bar{e}}^2 + 3 y_{\bar{u}}^2 + 3 y_{\bar{d}}^2 + \lambda_L^2 +  \lambda_E^2 + \lambda^2 \nonumber \\
&-&	 2 (\frac{9}{10} g_1^2 + \frac{3}{2} g_2^2 )] \nonumber
\end{eqnarray}
\begin{eqnarray}
\beta_{\lambda}&=&\frac{1}{16 \pi^2} \lambda [ 4 \lambda^2 + 6 \lambda_Q^2 + 3 \lambda_U^2 + 3 \lambda_D^2 + 2 \lambda_L^2 + \lambda_E^2 + 3 y_u^2 + 3 y_{\bar{u}}^2 + 3 y_d^2 + 3 y_{\bar{d}}^2 + y_e^2 + y_{\bar{e}}^2 + 3 y_t^2 \nonumber \\
&-&	2 (\frac{3}{20} g_1^2 + \frac{3}{4} g_2^2 )] \nonumber
\end{eqnarray}
\begin{eqnarray}
\beta_{\kappa}&=&\frac{1}{16 \pi^2} 3 \kappa [ \kappa^2 / 3 + 3 \lambda_Q^2 + 3 \lambda_U^2 + 3 \lambda_D^2 +  2 \lambda_L^2 + \lambda_E^2 + 3 \lambda^2] \nonumber
\end{eqnarray}
\begin{eqnarray}
\beta_{y_t}&=&\frac{1}{16 \pi^2} y_t [ 6 y_t^2 + 3 y_{\bar{d}}^2 + 3 y_u^2 + 3 y_{\bar{e}} + \lambda^2 \nonumber \\
&-&	 2 (\frac{13}{30} g_1^2 + \frac{3}{2} g_2^2 + \frac{8}{3} g_3^2 )]. \label{Yukawa_RG}
\end{eqnarray}
If we do not calculate complete $5+\overline{5}+10+\overline{10}$ RG-flows, we can just set the irrelevant yukawa coupling constants into 0.


\newpage

\begin{thebibliography}{99}
\bibitem{Primer} 
  For a review see S.~P.~Martin,
  Adv.\ Ser.\ Direct.\ High Energy Phys.\  {\bf 21}, 1 (2010)
  [hep-ph/9709356].



\bibitem{SUSY_GUT} 
P. Langacker, M.-X. Luo, Phys. Rev. D 44 (1991) 817; C. Giunti, C.W. Kim, U.W. Lee, Mod.
Phys. Lett. A 6 (1991) 1745; U. Amaldi, W. de Boer, H. Furstena
u, Phys. Lett. B 260 (1991)
447; J. Ellis, S. Kelley, D. Nanopolous, Phys. Lett. B 260 (19
91) 131

\bibitem{4th1}
For a review, see 
P.H. Frampton, P.Q. Hung and M. Sher, Phys. Rept. 330 (2000) 263.  

\bibitem{4th2}
For examples, see 
B. Holdom, Phys. Rev. Lett. 57 (1986) 2496 (Erratum-ibid 58 (1987) 177); 
W.S. Hou, R.S. Willey and A. Soni, Phys. Rev. Lett. 58 (1987) 1608 
[Erratum-ibid. 60 (1988) 2337];
M.S. Carena, H.E. Haber and C.E.M. Wagner, 
Nucl. Phys. B 472 (1996) 55;  
C.-S. Huang, W.-J. Huo and Y.-L. Wu, Phys. Rev. D 64 (2001) 016009;
Z. Murdock, S. Nandi and Z. Tavartkiladze, Phys. Lett. B 668 (2008) 303.  

\bibitem{4th3}
A.K. Grant and Z. Kakushadze, Phys. Lett. B 465 (1999) 108.  

\bibitem{4th4}
H.-J. He, N. Polonsky and S.-F. Su, Phys. Rev. D 64 (2001) 053004;
G.D. Kribs, T. Plehn, M. Spannowsky and T.M.P. Tait, 
Phys. Rev. D 76 (2007) 075016;
  N.~Chen and H.~J.~He,
  JHEP 1204, 062 (2012)
  [arXiv:1202.3072 [hep-ph]].

\bibitem{VL}
K.S. Babu, J.C. Pati and H. Stremnitzer, Phys.Lett.B 256, 206(1991); T. Moroi and Y. Okada, Mod.Phys.Lett.A 7, 187 (1992); Phys.Lett.B 295, 73 (1992); K.S. Babu and J.C. Pati, Phys.Lett.B 384, 140 (1996); M. Bastero-Gil and B. Brahmachari, Nucl.Phys.B 575, 35 (2000); Q. Shafi and Z. Tavartkiladze, Nucl.Phys.B 580, 83 (2000); K.S. Babu, I.
Gogoladze and C. Kolda, hep-ph/0410085; V. Barger, J. Jiang, P. Langacker and T.-J. Li, Int.J.Mod.Phys.A 22, 6203 (2007); K.S. Babu, I. Gogoladze, M.U. Rehman, Q. Shafi,Phys.Rev.D 78, 055017 (2008); P.W. Graham, A. Ismail, S. Rajendran, P. Saraswat, Phys.Rev.D 81, 055016 (2010).

\bibitem{Martin:2009bg} 
  S.~P.~Martin,
  Phys.\ Rev.\ D {\bf 81}, 035004 (2010)
  [arXiv:0910.2732 [hep-ph]].
  
\bibitem{Martin:2010dc} 
  S.~P.~Martin,
  Phys.\ Rev.\ D {\bf 82}, 055019 (2010)
  [arXiv:1006.4186 [hep-ph]].

\bibitem{Liu:2009cc} 
  C.~Liu,
  Phys.\ Rev.\ D {\bf 80}, 035004 (2009)
  [arXiv:0907.3011 [hep-ph]].

\bibitem{Liu_With_Lu} 
  C.~Liu and J.~S.~Lu,
  JHEP {\bf 1305}, 040 (2013)
  [arXiv:1305.0070 [hep-ph]].

\bibitem{Chang:2013eia} 
  X.~Chang, C.~Liu and Y.~L.~Tang,
  Phys.\ Rev.\ D {\bf 87}, no. 7, 075012 (2013)
  [arXiv:1303.7055 [hep-ph]].


\bibitem{TopSeeSaw}
  H.~J.~He, T.~M.~P.~Tait and C.~P.~Yuan,
  Phys.\ Rev.\ D {\bf 62}, 011702 (2000)
  [hep-ph/9911266];
    H.~J.~He, C.~T.~Hill and T.~M.~P.~Tait,
  Phys.\ Rev.\ D {\bf 65}, 055006 (2002)
  [hep-ph/0108041].
    X.~F.~Wang, C.~Du and H.~J.~He,
  Phys.\ Lett.\ B {\bf 723}, 314 (2013)
  [arXiv:1304.2257 [hep-ph]].
  
\bibitem{GMSB_GUT} 
  It is inferred from the method described in G.~F.~Giudice and R.~Rattazzi,
  Phys.\ Rept.\  {\bf 322}, 419 (1999)
  [hep-ph/9801271], see Page 9.
  
\bibitem{Elisabetta} 
  S.~Dawson and E.~Furlan,
  Phys.\ Rev.\ D {\bf 86}, 015021 (2012)
  [arXiv:1205.4733 [hep-ph]];
  S.~Dawson, E.~Furlan and I.~Lewis,
  Phys.\ Rev.\ D {\bf 87}, 014007 (2013)
  [arXiv:1210.6663 [hep-ph]];
  S.~Dawson and E.~Furlan,
  Phys.\ Rev.\ D {\bf 89}, 015012 (2014)
  [arXiv:1310.7593 [hep-ph]].


\bibitem{NMSSMInt} 
  For a review see U.~Ellwanger, C.~Hugonie and A.~M.~Teixeira,
  Phys.\ Rept.\  {\bf 496}, 1 (2010)
  [arXiv:0910.1785 [hep-ph]].


\bibitem{NMSSMVLAncester1} 
  M.~Masip, R.~Munoz-Tapia and A.~Pomarol,
  Phys.\ Rev.\ D {\bf 57}, R5340 (1998)
  [hep-ph/9801437].
  
\bibitem{NMSSMVLAncester2} 
  J.~R.~Espinosa and M.~Quiros,
  Phys.\ Rev.\ Lett.\  {\bf 81}, 516 (1998)
  [hep-ph/9804235].

\bibitem{NMSSMVLAncester3} 
  Y.~Daikoku and D.~Suematsu,
  Prog.\ Theor.\ Phys.\  {\bf 104}, 827 (2000)
  [hep-ph/0003206].

\bibitem{HeHongjianS} 
  H.~J.~He and Z.~Z.~Xianyu,
  arXiv:1405.7331 [hep-ph].

\bibitem{Earlier} 
  Y.~Okada and L.~Panizzi,
  Adv.\ High Energy Phys.\  {\bf 2013}, 364936 (2013)
  [arXiv:1207.5607 [hep-ph]].
  
\bibitem{CMS_Limit} 
  S.~Chatrchyan {\it et al.}  [CMS Collaboration],
  Phys.\ Lett.\ B {\bf 729}, 149 (2014)
  [arXiv:1311.7667 [hep-ex]].


\bibitem{LotsofMistakes} 
  W.~Wang, J.~M.~Yang and L.~L.~You,
  JHEP {\bf 1307}, 158 (2013)
  [arXiv:1303.6465 [hep-ph]].
  
\bibitem{PDG}
J. Beringer et al. (Particle Data Group), Phys. Rev. D86, 010001 (2012)
and 2013 partial update for the 2014 edition. 

\bibitem{SUSYRun_1} 
D.R.T. Jones, Nucl.Phys.B 87, 127 (1975). D.R.T. Jones and L. Mezincescu, Phys.Lett.B 136, 242 (1984). P.C. West, Phys.Lett.B 137, 371 (1984). A. Parkes and P.C. West, Phys. Lett. B 138, 99 (1984).

\bibitem{SUSYRUN_2}
 S.P. Martin and M.T. Vaughn, Phys.Lett.B 318, 331 (1993) [hep-ph/9308222], Phys.Rev.D 50, 2282 (1994) [Erratumibid. D 78, 039903 (2008)] [hep-ph/9311340]. Y. Yamada, Phys.Rev.D 50, 3537 (1994) [hep-ph/9401241]. I. Jack and D.R.T. Jones, Phys.Lett.B 333, 372 (1994) [hep-ph/9405233]. I. Jack et al, Phys.Rev.D 50, 5481 (1994) [hep-ph/9407291]

\end{thebibliography}
\end{document}